\documentclass{PoS}
\usepackage{subfigure}
\usepackage[intlimits]{amsmath}
\usepackage{caption}
\def\slashed#1{\kern+0.1em /\kern-0.65em #1}

\title{Calculating the $K_L-K_S$ mass difference and $\epsilon_K$ to sub-percent accuracy}

\ShortTitle{Calculating the $K_L-K_S$ mass difference and $\epsilon_K$ to sub-percent accuracy}

\newcommand\bnl{Brookhaven National Laboratory, Upton, NY 11973, USA}
\newcommand\cu{Physics Department, Columbia University, New York,
      NY 10027, USA}
\newcommand\riken{RIKEN-BNL Research Center, Brookhaven National
      Laboratory, Upton, NY 11973, USA}
\newcommand\soton{School of Physics and Astronomy, University of
  Southampton,  Southampton SO17 1BJ, UK}

\author{\speaker{Norman Christ}\thanks{This work was supported in part by US DOE grant DE-FG02-92ER40699.}\\ \cu \\ E-mail: \email{nhc@phys.columbia.edu}}
\author{Taku Izubuchi\\ \bnl \riken \\ E-mail: \email{izubuchi@quark.phy.bnl.gov}}
\author{Christopher T. Sachrajda\\ \soton \\ E-mail: \email{cts@soton.ac.uk}}
\author{Amarjit Soni\\ \bnl \\ E-mail: \email{adlersoni@gmail.com}}
\author{Jianglei Yu\\ \cu \\ E-mail: \email{physhark@gmail.com}} 
\author{RBC and UKQCD Collaborations}

\abstract{The real and imaginary parts of the $K_L-K_S$ mixing 
matrix receive contributions from all three charge-2/3 
quarks: up, charm and top.  These give both short- 
and long-distance contributions which are accessible 
through a combination of perturbative and lattice 
methods.  We will discuss a strategy to compute both 
the mass difference, $\Delta M_K$ and $\epsilon_K$ to 
sub-percent accuracy, looking in detail at the 
contributions from each of the three CKM matrix element 
products $V_{id}^*V_{is}$ for $i=u, c$ and $t$ as described
in Ref.~\cite{Christ:2012se}.
}

\FullConference{31st International Symposium on Lattice Field Theory LATTICE 2013\\
		 July 29 - August 3, 2013\\
		 Mainz, Germany}

\begin{document}

\section{Introduction}

The mixing between the $K^0$ meson and it anti-particle $\overline{K}^0$ is both highly sensitive to physics that lies outside of the standard model and also has been measured experimentally to an impressively high accuracy.  In fact, this system is often presented as a textbook example of quantum mechanical mixing of two unstable states.  In the non-covariant, Wigner-Weisskopf treatment the $K^0 - \overline{K}^0$ system is described by two complex, time-dependent amplitudes $K^0(t)$ and $\overline{K}^0(t)$ which, when arranged as a two-component vector obey:
\begin{equation}
i\frac{d}{dt}\left(\begin{array}{c} K^0 \\ \overline{K}^0 \end{array}\right) 
= \left\{ \left( \begin{array}{cc} M_{00} & M_{0\overline{0}} \\
                                 M_{\overline{0}0} &M_{\overline{0}\overline{0}}
                           \end{array} \right) 
- \frac{i}{2} \left( \begin{array}{cc} \Gamma_{00} & \Gamma_{0\overline{0}} \\
                                 \Gamma_{\overline{0}0} &\Gamma_{\overline{0}\overline{0}}
                           \end{array} \right)\right\} 
\left(\begin{array}{c} K^0 \\ \overline{K}^0 \end{array}\right)
\end{equation}
where the matrix $\Gamma_{ij}$ can be constructed from energy-conserving,
$K\to\pi\pi$ matrix elements:
\begin{equation}
\Gamma_{ij} = 2\pi \sum_\alpha \int_{2m_\pi}^\infty d E \langle i |H_W|\alpha(E)\rangle
                                \langle \alpha(E)|H_W|j\rangle \delta(E-m_K)  
\label{eq:Gamma}
\end{equation}
while the ``mass matrix'' $M_{ij}$ contains a sum over all intermediate states energies and uses the principal part to resolve singularities for states with an energy $E$ equal to $M_K$:
\begin{equation}
M_{ij} = \sum_\alpha \mathcal{P} \int_{m_\pi}^\infty d E \frac{\langle i |H_W|\alpha(E)\rangle
                                \langle \alpha(E)|H_W|j\rangle}{m_K - E}.
\label{eq:M}
\end{equation}
While of great interest in their own right, the $K\to\pi\pi$ matrix elements in Eq.~\eqref{eq:Gamma} are on-shell, $\langle\pi\pi|H_W|K\rangle$ matrix elements, which can be computed using lattice methods. Here we focus on the truly second order quantity $M_{ij}$ given in Eq.~\eqref{eq:M}.

The hermiticity of the weak operator $H_W$ and CPT invariance imply that the two diagonal elements, $M_{00}$ and $M_{\overline{0}\overline{0}}$ are real and equal.  They represent a common shift in the masses of the two $K^0-\overline{K}^0$ decaying eigenstates, do not appear to be of great interest and also are not the target of the present discussion.   The off-diagonal element $M_{0\overline{0}} = M_{\overline{0}0}^*$ is of fundamental importance in particle physics.  The real part of $M_{0\overline{0}}$ gives the $3.486 \times 10^{-12}$ MeV $K_L-K_S$ mass difference $\Delta M_K$ while the imaginary part makes the largest contribution to the indirect CP violation parameter $\epsilon_K$:
\begin{equation}
\epsilon_K = \frac{i}{2}\left\{\frac{{\rm Im} M_{0\overline{0}}
                            -\frac{i}{2}{\rm Im} \Gamma_{0\overline{0}} }
                                  {{\rm Re} M_{0\overline{0}}
                            -\frac{i}{2}{\rm Re} \Gamma_{0\overline{0}} }\right\}
                 + i\frac{{\rm Im}A_0}{{\rm Re}A_0}, 
\quad
\Delta M_K=m_{K_S}-m_{K_L} = 2 {\rm Re}\{M_{0\overline{0}}\}.
\end{equation}

In the standard model the real and imaginary parts of $M_{0\overline{0}}$ receive their largest contributions from two quite different sources.  Since no small CP violating terms are needed, Re$(M_{0\overline{0}})$ is dominated by states which couple most strongly to the $K^0$ and $\overline{K}^0$, those containing the up and charm quarks and by energies at or below the charm quark mass.  The real part of $(M_{0\overline{0}})$ has proven an challenging quantity to compute since these charm-scale energies require a non-perturbative treatment~\cite{Brod:2011ty} which has only recently become available~\cite{Yu:2011np, Yu:2012nx, Christ:2012se}.  In contrast, Im$(M_{0\overline{0}})$ requires the presence of CP violation and therefore in the standard model the participation of all three charge-$\frac{2}{3}e$ quarks.  The top quark must contribute through loop effects which are dominated by energies on the order of the top quark mass $m_t$.  Electroweak and QCD perturbation theory can be used to represent this dominant top quark contribution as a known Wilson coefficient times an effective, low-energy, four-quark operator $O_{LL}$,
\begin{equation}
O_{LL} = \overline{s}\gamma^\mu(1-\gamma^5)d\overline{s}\gamma^\mu(1-\gamma^5)d,
\label{eq:OLL}
\end{equation}
whose $K^0-\overline{K}^0$ matrix element gives $B_K$, the target of lattice calculations for twenty five years.

In this talk, we would like to go beyond these leading contributions to the complex mixing amplitude $M_{0\overline{0}}$ and discuss sub-leading effects which will give corrections on the few percent scale and therefore must be well understood if we are to compute $\Delta M_K$ and $\epsilon_K$ to sub-percent accuracy in the standard model.

\section{General framework}

\begin{figure}[ht]
\centering
\subfigure[\label{fig:box}]{\includegraphics[width=0.42\textwidth]{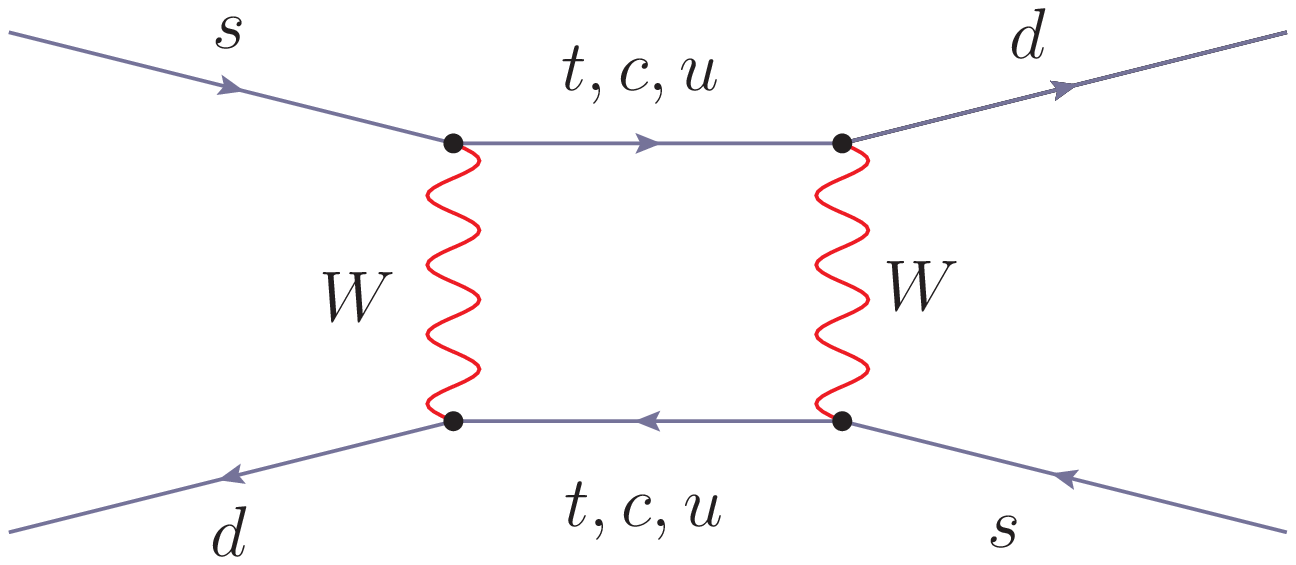}}
\subfigure[\label{fig:disconnected}]{\includegraphics[width=0.35\textwidth]{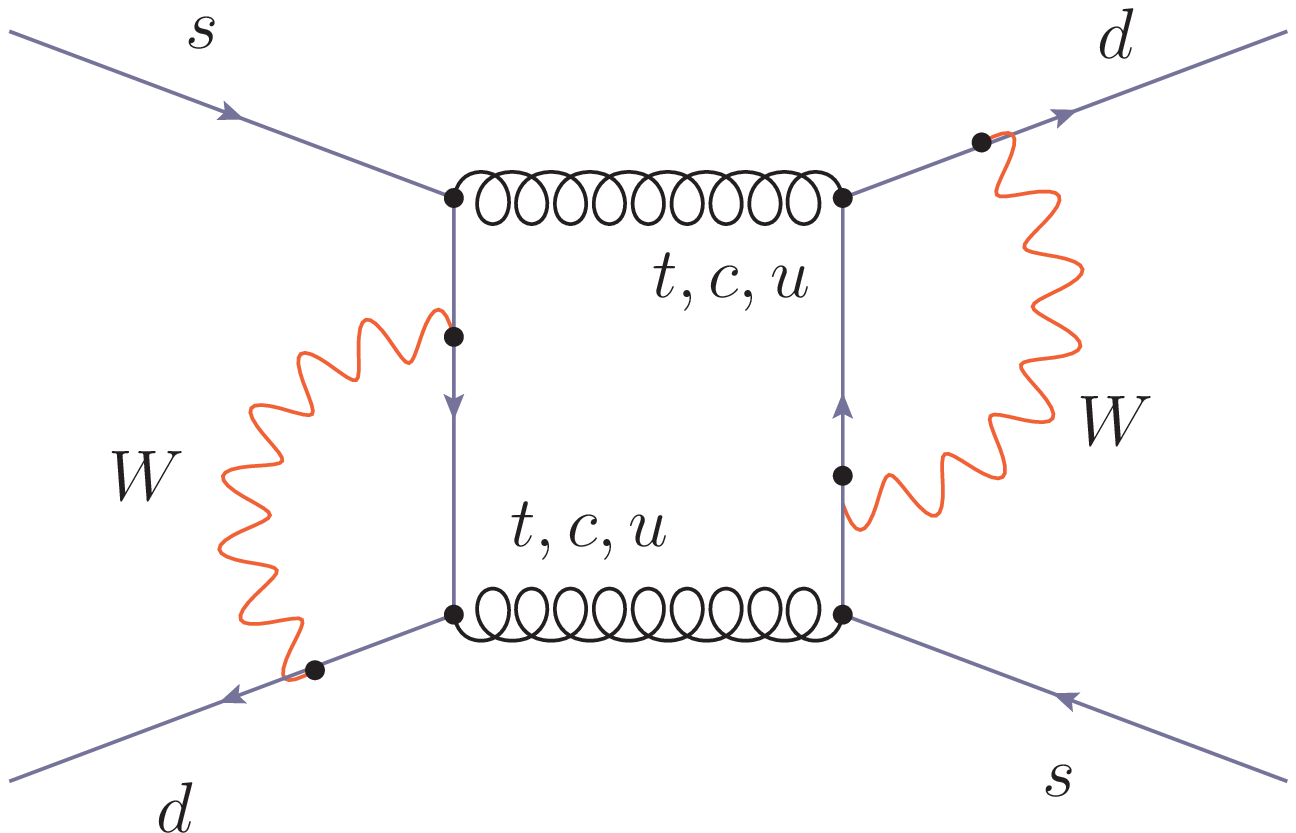}}
\caption{Sample diagrams representing the two classes of graphs which enter any standard model calculation of $M_{0\overline{0}}$. In all diagrams in the standard model contributing to $M_{0\overline{0}}$ one can follow two quark lines which enter the diagram as strange quarks and exit as down quarks.  Each quark line has two $W^\pm$ couplings, one which changes strangeness and the other which does not.  The two classes of diagrams are distinguished by the manner in which these $W^\pm$ verticies are connected: either the $W$ lines connect (a) each quark line to the other or (b) each quark line to itself.  } 
\label{fig:general}
\end{figure}

While Eq.~\eqref{eq:M} provides a familiar expression representing much of the physics associated with $(M_{0\overline{0}})$, the presence of the $\Delta S=1$, effective weak operator $H_W$ implies a long distance approximation which cannot be made for all contributions to  $(M_{0\overline{0}})$ in the standard model.   Instead we must examine general Feynman amplitudes which connect  $K^0$ and $\overline{K}^0$ states and involve the exchange of two $W^\pm$ bosons.  These are of the two types shown in Fig.~\ref{fig:general}.  We refer to those in Fig.~\ref{fig:box} as ``box diagrams'' and those in
Fig.~\ref{fig:disconnected} as "disconnected diagrams''.

The contributions of the various quark flavors to $\Delta M_K$ and $\epsilon_K$ are strongly affected by the quark masses, the Glashow, Iliopoulos and Maiani (GIM) cancellation and the sizes of the real and imaginary parts of the corresponding CKM matrix elements.  The combined effects of these three aspects are usually represented in a sub-diagram with intermediate up, charm and top quarks, such as that shown in Fig.~\ref{fig:GIM}, by performing a subtraction of the identical amplitude in which the varying quark mass in the intermediate quark propagator has been replaced by that of the up quark~\cite{Inami:1980fz}.  Such a sum over the three flavors but with a fixed quark mass will vanish exactly because of the GIM mechanism (or the orthogonality of the strange and down columns of the CKM matrix).  

This conventional approach has two disadvantages for our purposes.  First it creates a term which combines top and up quark propagators involving both very long and very short distances.  Second it results in a charm and up quark combination which contains a CP violating phase.  This apparent long-distance contribution to $\epsilon_K$ is, of course, not actually present and is canceled by a similar phase in the top-up contribution.  Instead, we choose to subtract the term in which the common quark mass is taken to be that of the charm quark.   Thus, if no gluons couple to this intermediate quark line we would use:
\begin{eqnarray}
\sum_{i=u,c,t}&&\left\{ V_{i,d}^* \frac{\slashed{p}}{p^2+m_i^2} V_{i,s}
                    - V_{i,d}^* \frac{\slashed{p}}{p^2+m_c^2} V_{i,s}
              \right\}  \label{eq:GIM} \\
              && = \lambda_t\left\{\frac{\slashed{p}}{p^2+m_t^2}
                                                  -\frac{\slashed{p}}{p^2+m_c^2}
                                        \right\}
                 + \lambda_u\left\{\frac{\slashed{p}}{p^2+m_u^2}
                                                  -\frac{\slashed{p}}{p^2+m_c^2}
                                        \right\}, \nonumber
\end{eqnarray}
where $\lambda_i = V_{i,d}^*V_{i,s}$.  In this expression the GIM cancellation is explicit, the top quark is combined only with the heavier charm quark and there is no CP violation associated with the up-quark.  This same approach can be taken for a general diagram using the full quark propagators, including interactions with the gluon field.  

Since two quark lines appear in each diagram, we can distinguish a total of six different contributions to $\Delta M_K$ and $\epsilon_K$: those from the two types of diagram with the three possible products $\lambda_u\lambda_u$, $\lambda_u\lambda_t$ and $\lambda_t\lambda_t$.

\begin{center}
\begin{minipage}{0.45\textwidth}
\centering
\includegraphics[width=0.9\textwidth]{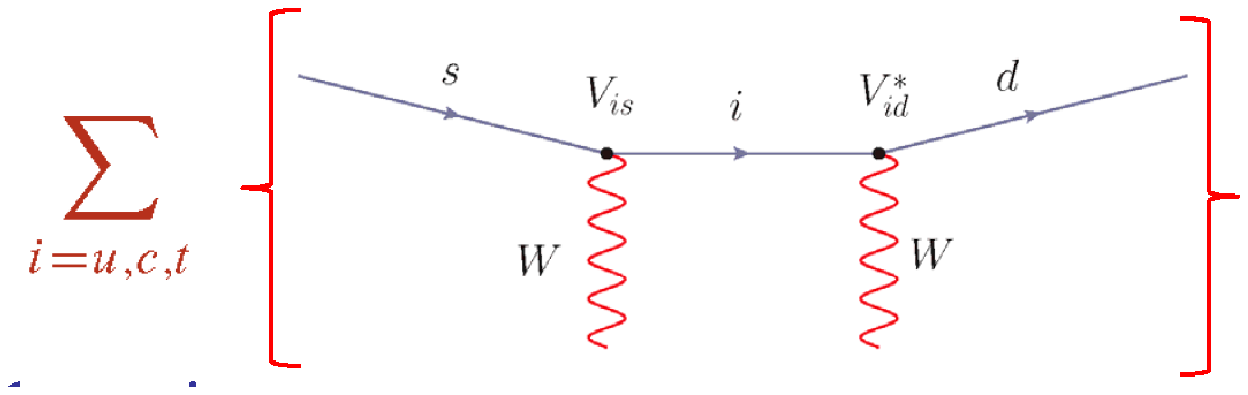} 
\captionof{figure}{A Feynman diagram showing the three $2/3e$ quarks as intermediate states in one of the two quark lines which pass through a standard model, $\Delta S=2$ diagram.  Such a diagram can also include any number of gluon lines which are suppressed here.
\label{fig:GIM}}
\end{minipage} \quad
\begin{minipage}{0.45\textwidth}
\centering
\includegraphics[width=0.87\textwidth]{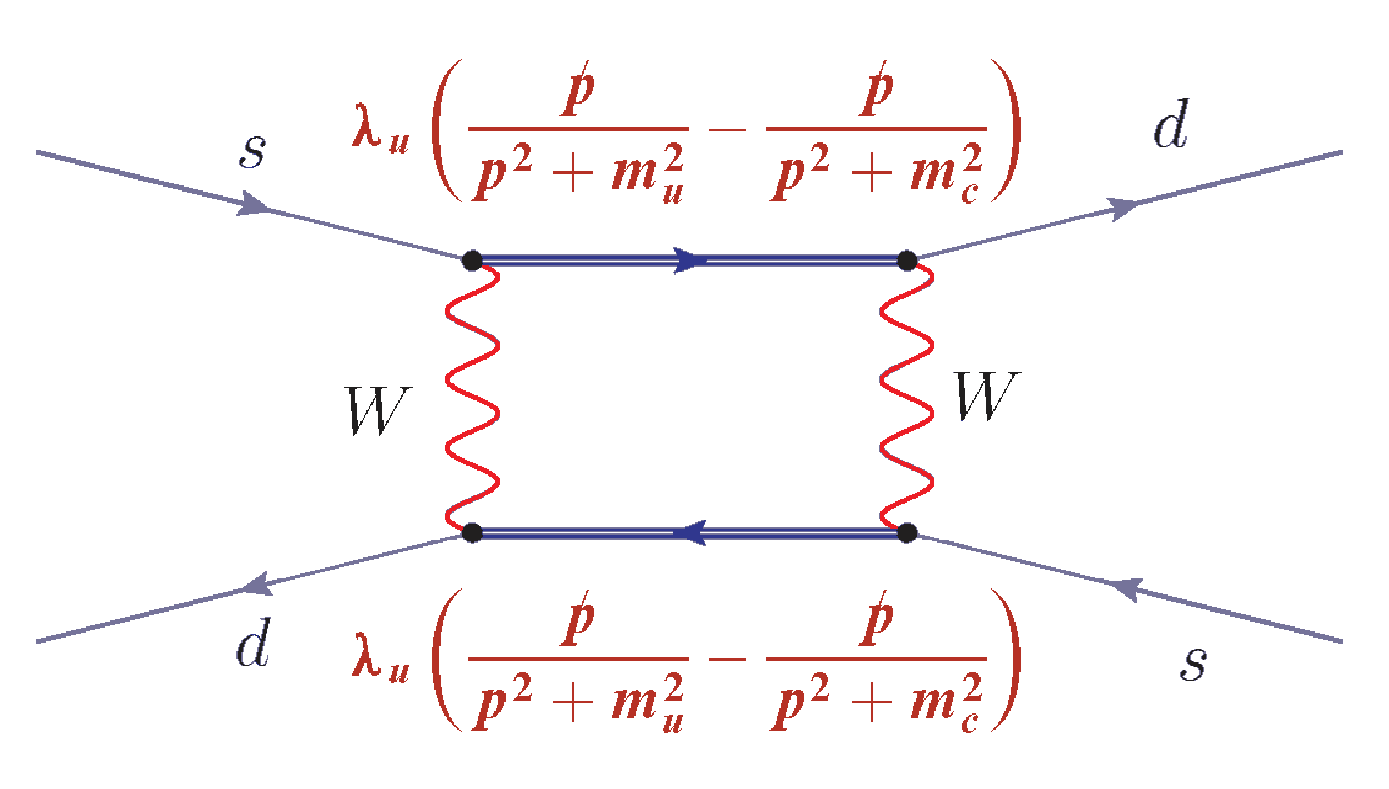}
\captionof{figure}{A Feynman diagram showing the contribution to $M_{0\overline{0}}$ of a box diagram  where both quark lines contain a $u-c$ subtraction.
\label{fig:box_uu}}
\end{minipage} 
\end{center}

\section{Six contributions to $\Delta M_K$ and $\epsilon_K$}
We will now discuss each of these six contributions in turn, keeping in mind the approximate experimental values for the three quantities $\lambda_u$, $\lambda_c$ and $\lambda_t$ 
\begin{equation}
\lambda_u = 0.22, \quad \lambda_c = -0.22 + 1.34 \times 10^{-4} i \quad\mbox{and}\quad
 \lambda_t =  3.2 \times 10^{-4} -  1.34 \times 10^{-4} i
\end{equation}
and the large mass ratio $(m_t/m_c)^2 = 2.1 \times 10^4$.

\underline{\bf $\mathbf{uu}$ box diagram.}  This contribution comes from the box diagram in Fig.~\ref{fig:box} in which the sum over up, charm and top quark propagators in each of the quark lines is replaced by the difference of an up and charm propagator and the four CKM matrix elements can be written $\lambda_u\lambda_u$.   The result is shown in Fig.~\ref{fig:box_uu}.  Since $\lambda_u$ is real this term contributes only to $\Delta M_K$.  The left-handed structure of the weak vertices implies that the masses of the intermediate charge-2/3 quark propagators will only enter quadratically so the GIM subtraction that occurs in each line will result in a subtracted propagator behaving at large momentum as $\slashed{p}/p^4$.  Having two such propagators within the box diagram insures convergence even when the $W$-exchange is treated as occurring at a point.  The result is an amplitude in which the intermediate momenta are on the scale of the charm mass or smaller and which has an over-all size of order $G_F^2 m_c^2\lambda_u^2$, as given in Tab.~\ref{tab:sizes} below.

Such an amplitude can be explicitly evaluated in lattice QCD provided the lattice spacing $a$ is sufficiently small that the charm quark can be accurately included in the calculation.  A first calculation of this dominant contribution to $\Delta M_K$ has been carried out~\cite{Christ:2012se} and a full calculation, including all diagrams, is nearly complete~\cite{Yu:2013qfa}.  (Note, the $O(m_c a)^2 \approx 0.4$ errors must still be reduced.) 

\underline{\bf $\mathbf{tt}$ box diagram.}  Here the large quark mass implies that this contribution to $M_{0\overline{0}}$ will be dominated by momenta on the scale of $m_t$.  As a result this can be  accurately represented as the operator $O_{LL}$ given in Eq.~\eqref{eq:OLL} multiplied by a Wilson coefficient which can be reliably determined using electroweak and QCD perturbation theory.  The  $K^0-\overline{K}^0$  matrix element of $O_{LL}$, which determines the parameter $B_K$, is now accurately computed using lattice methods making this $tt$ contribution very well known.  As can be seen in Tab.~\ref{tab:sizes} this contribution dominates $\epsilon_K$ and provides an $\approx 4\%$ correction to the dominant $uu$ contribution to $\Delta M_K$.  Corrections to this approximation of large momentum dominance will be quite small, $\mathcal{O}\bigl((m_c/m_t)^2\bigr)$.

\begin{figure}
\centering
\includegraphics[width=0.70\textwidth]{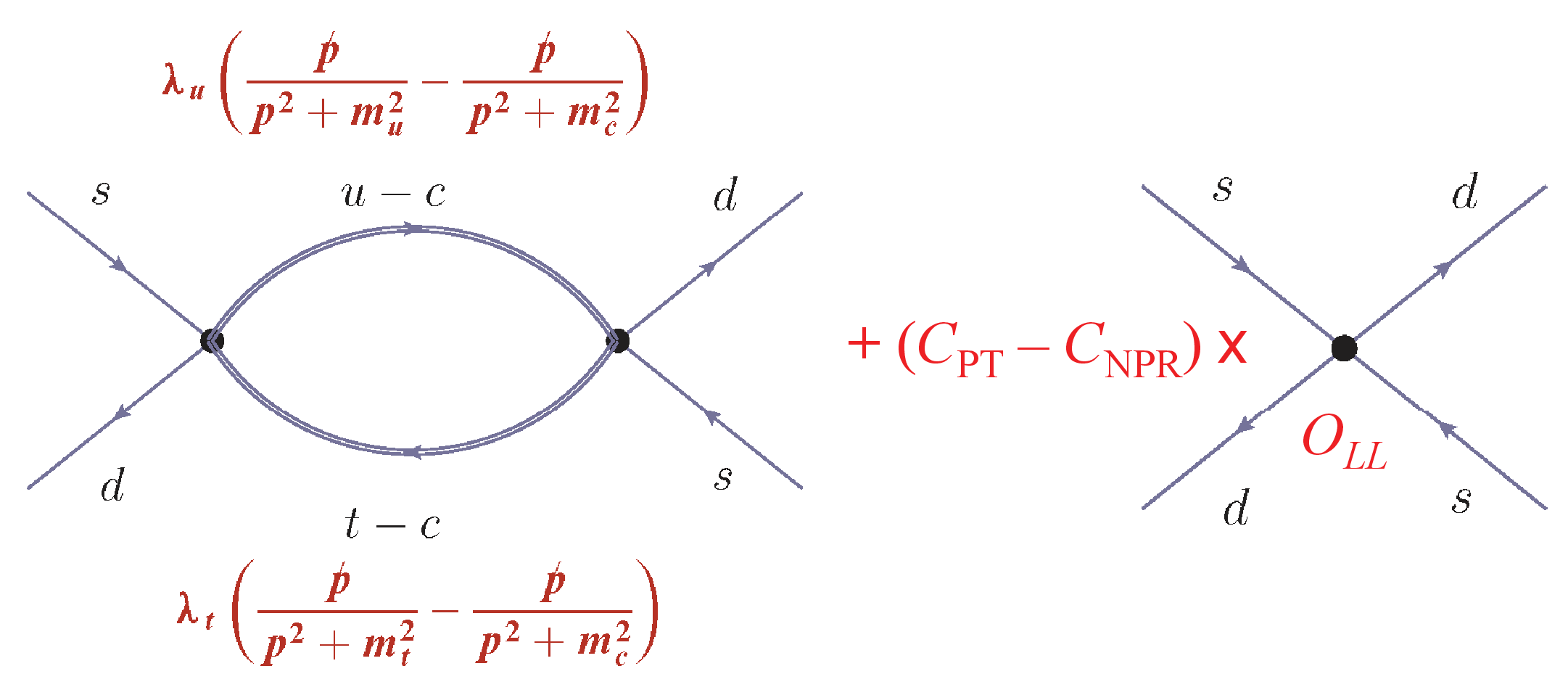} 
\caption{A Feynman diagram showing the contribution to $M_{0\overline{0}}$ of the box diagram where one quark line contains a $u - c$, and the other a $t-c$ subtraction.  The contribution from the top quark piece can be obtained from perturbation theory and a lattice evaluation of $\langle  K^0|O_{LL}|\overline{K}^0\rangle$. The charm contribution yields a $\log(1/m_c a)$ contribution which must be replaced by the physical $\log(m_W/m_c)$ result.  This can be done by the $O_{LL}$ operator subtraction given by the term on the right, with $C_\mathrm{NPR}$ the non-perturbatively determined lattice, and $C_\mathrm{PT}$ the perturbative continuum, short distance parts.
\label{fig:box_ut}}
\end{figure}

\underline{\bf $\mathbf{ut}$ box diagram.}  For this case one quark line involves a $u-c$ subtraction which falls as $\slashed{p}/p^4$ for $p > m_c$.  However, the second quark line contains the difference of top and charm quark propagators which are sufficiently different that they must be treated separately.  For momenta below the scale of the $W$ mass, the top quark propagator will behave as $\slashed{p}/m_t^2$ and the combined quark propagators with behave as $\slashed{p}^2/p^4$, resulting an integral which will grow quadratically until the energy scale of $m_W$ is reached.  Just as in the case of the $tt$ piece this term can be accurately represented as a known, perturbatively-determined coefficient multiplied by the operator $O_{LL}$ with corrections suppressed by $(m_c/m_W)^2$.   The term involving the charm quark propagator contains an additional factor of $1/p^2$ which results in a $\log(m_W/m_c)$ when the integral is performed.  If computed directly using lattice QCD, we instead expect a result containing $\log(1/m_c a)$, where the physical $m_W$ cut-off to the integration has been replaced by the lattice cut-off.

Never-the-less this quantity can be accurately computed using a combination of perturbative and lattice methods by combining two terms, as in Fig.~\ref{fig:box_ut}.  The presence of the lattice cut-off implies that the large-momentum contribution to this box graph (which will appear as a four-quark coupling at the scale of the QCD) involves the factor $\log(1/m_c a)$ instead of the correct short-distance expression $\propto \log(m_W/m_c)$.  This short distance part can be isolated in a lattice calculation by adopting a Rome-Southampton approach and evaluating the four external quark lines at large, non-exceptional momenta of scale $\mu \gg \Lambda_\mathrm{{QCD}}$.  The resulting amplitude will be infra-red safe, reflecting only momenta on the order of $\mu$.  Because of the size of the lattice cut-off this quantity will not have its physical value.  However, this specific, gauge-fixed quantity can be reliably computed in perturbation theory and the subtraction suggested in Fig.~\ref{fig:box_ut} performed to replace the lattice value of this off-shell gauge-fixed Green's function by it physical value.  This process changes the single, incorrectly-determined part of the lattice result to its physical value.  The subtracted lattice result then obeys a single condition which guarantees that it contains the correct short-distance part. Such a subtraction was successfully implemented in Ref.~\cite{Yu:2011np} and explained in greater detail in Ref.~\cite{Christ:2012se}.

\underline{\bf $\mathbf{uu}$ disconnected diagram.}  This contribution is illustrated in Fig.~\ref{fig:disconnected} when the difference of the full up minus charm quark propagators is inserted between the two $W^\pm$ vertices for both quark lines.  This double subtraction results in a convergent amplitude even when each $W$ exchange is treated as occurring at a point.  The result is real and contributes to $\Delta M_K$ a quantity similar in size to that coming from the $uu$ box diagram.  In fact, it appears that this OZI-suppressed contribution is large, of the same size as that coming from the box diagram~\cite{Yu:2013qfa}. 

\underline{\bf $\mathbf{tt}$ disconnected diagram.}  Just as in the case of the $tt$ box diagram, this term is dominated by momenta on the order of  $m_W$ or larger and can be reliably computed as the product of a perturbation theory Wilson coefficient and the hadronic matrix element $\langle  K^0|O_{LL}|\overline{K}^0\rangle$.  As suggested in Tab.~\ref{tab:sizes} this contributes on the $10^{-3}$ level to $\Delta M_K$ and may be a standard $\approx 10^{-2}$-level NNLO contribution to the perturbative Wilson coefficient that determines $\epsilon_K$.

\underline{\bf $\mathbf{ut}$ disconnected diagram.}  While ultimately accessible to a combination of lattice and perturbative methods, this is the most complex of the six cases.  Referring to Fig.~\ref{fig:disconnected}, it is natural to treat the amplitude as two factors joined by two or more gluon lines.  One factor contains the difference of up and charm quark propagators, a four-quark vertex formed by contracting the $W$ propagator to a point and at least one attached gluon line. Given the requirements of QCD gauge invariance, this portion of the diagram will be convergent and can be evaluated in a lattice calculation which includes the charm quark.  The second quark line involves a top and a charm quark propagator which, as above, should be treated separately.  For the case of top, subgraphs containing the one $W$ loop, external down and strange quark lines and a single gluon will be dominated by momenta on the order of $m_t$ and can be represented in this calculation by a gluonic penguin contribution:  If the quark line to which the gluon couples is also included, this will appear to be an insertion of a combination of the four, standard, gluonic penguin operators with coefficients of order $\lambda_t/m_W^2$ which can be computed in perturbation theory.  

The contribution from the charm  quark propagator requires more work.  When the $W$ propagator is treated as a point and gauge invariance incorporated, similar sub-diagrams with up, strange and gluon external lines will contain a logarithmically growing momentum integration cut off at the $m_W$ scale.  Thus, as for the charm quark contribution to the $ut$ box diagram, this must be evaluated using subtracted lattice methods to capture the correct long-distance contribution but to replace the short distance part by the correct $\log(m_W/m_c)$ term.  Finally, once these two $u-c$ and $t-c$ factors have been sorted out, their combination will again lead to logarithmically divergent subgraphs which will also require a subtracted lattice QCD evaluation.  A double subtraction of this sort has not yet been attempted but appears to be well-defined and should be possible.

\section{Summary and Conclusion}

The largest contribution to $\epsilon_K$ is now a standard result from lattice QCD with the largest errors arising from imperfectly known standard model parameters, presently $V_{cb}$.  While less developed, one expects lattice results for $\Delta M_K$, accurate at the level of 10\% in the next 2-3 years .  Here, following Appendix A of Ref.~\cite{Christ:2012se}, we have argued that the sub-leading contributions to $\epsilon_K$ and $\Delta M_K$, summarized in Tab.~\ref{tab:sizes} should also be accessible to lattice methods, allowing the theoretical errors on the standard model predictions for these quantities to be eventually reduced below the 1\% level.  At this level of accuracy, the effects of electromagnetism and the iso-spin breaking, up-down quark mass difference need to be included, a topic outside of the discussion presented here.

\begin{table}
\center
\begin{tabular}{c|ccc}
quarks          & $M_{0\overline{0}}$                 & Re($M_{0\overline{0}}$)
                                                                                                             & Im($M_{0\overline{0}}$) \\
\hline\hline
$(u-c)(u-c)$ & $\lambda_u^2 (m_c/m_W)^2$ & $1.1\times 10^{-5}$ & $0$ \\
$(t-c)(t-c)$   & $\lambda_t^2 (m_t/m_W)^2$   & $4.0\times 10^{-7}$ & $4.1 \times 10^{-7}$ \\
$(u-c)(t-c)$  & $\lambda_u\lambda_t(m_c/m_W)^2$
                                                                         & $1.6\times 10^{-8}$ & $6.6 \times 10^{-9}$
\end{tabular}
\caption{Estimates of the sizes of the various contribution to $M_{0\overline{0}}$ from the three products of the two different quark combinations which result from the GIM cancellation implemented as in Eq.~(2.1).  We do not attempt to distinguish the relative sizes coming from the box and disconnected diagrams.}
\label{tab:sizes}
\end{table}

\providecommand{\href}[2]{#2}\begingroup\raggedright\endgroup

\end{document}